\documentclass[epj]{svjour}
\usepackage{epsfig}
\usepackage{graphics}
\def\nA{nucleon-nucleus\ }
\def\pA{proton-nucleus\ }

\begin{document}
\title{Isovector deformation and its link to the neutron shell closure}
\author{Dao T. Khoa}
\institute{Institute for Nuclear Science {\rm \&} Technique, VAEC, P.O. Box
 5T-160, Nghia Do, Hanoi, Vietnam \\ \email{khoa@vaec.gov.vn}}
\date{Received: date / Revised version: date}
%
\abstract{DWBA analysis of the inelastic $^{30-40}$S$(p,p')$ and
$^{18-22}$O$(p,p')$ scattering data measured in the inverse kinematics has been
performed to determine the isoscalar ($\delta_0$) and isovector ($\delta_1$)
deformation lengths of the 2$^+_1$ excitations in the Sulfur and Oxygen isotopes
using a compact folding approach. A systematic $N$-dependence of $\delta_0$ and
$\delta_1$ has been established which shows a link between $\delta_1$ and the
neutron-shell closure. Strong isovector deformations were found in several
cases, e.g., the 2$^+_1$ state in $^{20}$O where $\delta_1$ is nearly three
times larger than $\delta_0$. These results confirm the relation
$\delta_1>\delta_0$ anticipated from the core polarization by the valence
neutrons in the open-shell (neutron rich) nuclei. The effect of neutron shell
closure at $N=14$ or 16 has been discussed based on the folding model analysis
of the inelastic $^{22}$O+$p$ scattering data at 46.6 MeV/u measured recently at
GANIL.
\PACS{ 25.40.Ep, 21.10.Re, 24.10.Eq, 24.10.Ht}
} 
\maketitle
The neutron and proton contributions to the structure of the lowest 2$^+$
excited states are known to be quite different in the neutron-rich nuclei due,
in particular, to a strong polarization of the core by valence neutrons
\cite{Be83}. In the distorted-wave Born approximation (DWBA) for the inelastic
hadron scattering, the different neutron and proton contributions to the nuclear
excitation are explicitly determined by the isospin dependence of the inelastic
form factor (FF).

In general, the isospin-dependent nucleon optical potential (OP) can be written
in terms of the isoscalar (IS) and isovector (IV) components \cite{La62} as
\begin{equation}
 U(R)=U_0(R)\pm\varepsilon U_1(R), \ \varepsilon=(N-Z)/A, \label{e1}
\end{equation}
where the + sign pertains to incident neutron and - sign to incident proton.
While the strength of the Lane potential $U_1$ has been studied since a long
time \cite{La62} in the $(p,p)$ and $(n,n)$ elastic scattering and $(p,n)$
reaction studies, very few attempts were made to study the isospin dependence of
the inelastic FF. Within a collective-model prescription, the inelastic FF for
the \nA scattering is obtained by ``deforming" the OP (\ref{e1}) with scaling
factors $\delta$ known as the nuclear deformation lengths
\begin{equation}
 F(R)=\delta\frac{dU(R)}{dR}=\delta_0\frac{dU_0(R)}{dR}\pm\varepsilon\
 \delta_1\frac{dU_1(R)}{dR}. \label{e2}
\end{equation}
The explicit knowledge of the isoscalar ($\delta_0$) and, especially, isovector
($\delta_1$) deformation lengths would give us vital information on the
structure of the nuclear excitation under study. There are only two types of
experiment that might allow one to determine the IV deformation length
$\delta_1$ based on the prescription (\ref{e2}):

i) $(p,n)$ reaction leading to the \emph{excited} isobar analog state. It was
shown \cite{Fi79}, however, that the two-step mechanism usually dominates this
process and the calculated DWBA cross sections are not sensitive to $\delta_1$.

ii) Another way is to extract $\delta_0$ and $\delta_1$ from the $(p,p')$ and
$(n,n')$ inelastic scattering measured at the same incident energy and exciting
the same target state \cite{Fi79,Gr80}. Such double measurements are presently
not feasible with the beams of unstable nuclei.

We have recently suggested a compact folding method \cite{Kh03} to determine
$\delta_{0(1)}$ based on the DWBA analysis of the $(p,p')$ data only. In this
approach, instead of deforming the OP, we build up the proton and neutron
transition densities of a $2^{\lambda}$-pole excitation ($\lambda\ge 2$) by
using the Bohr-Mottelson prescription \cite{Bo75} separately for protons and
neutrons
\begin{equation}
 \rho^\tau_\lambda(r)=-\delta_\tau\frac{d\rho^\tau_{\rm g.s.}(r)}{dr},\ {\rm with}
 \ \tau=p,n. \label{e3}
\end{equation}
Here $\rho^\tau_{\rm g.s.}(r)$ are the proton and neutron ground-state (g.s.)
densities and $\delta_\tau$ are the corresponding deformation lengths. Given the
explicit proton and neutron transition densities, one can obtain from the
folding model \cite{Kh02} the inelastic \pA FF in terms of the IS and IV parts
as
\begin{equation}
 F(R)=F_0(R)-\varepsilon F_1(R), \label{e4}
\end{equation}
where $F_0(R)$ and $F_1(R)$ are determined \cite{Kh02} from the sum
($\rho^n_\lambda+\rho^p_\lambda$) and difference
($\rho^n_\lambda-\rho^p_\lambda$) of the neutron and proton transition densities
(\ref{e3}), respectively. One can see that $F_1(R)$ is just the prototype of the
Lane potential in the inelastic nucleon scattering. It is natural to represent
the IS and IV parts of the nuclear transition density as
\begin{equation}
 \rho^{0(1)}_{\lambda}(r)=\rho^n_{\lambda}(r)\pm\rho^p_{\lambda}(r).
 \label{e5}
\end{equation}
On the other hand, $\rho^{0(1)}_{\lambda}(r)$ can be obtained using the same
Bohr-Mottelson method, by deforming the IS and IV parts of the g.s. density
\begin{equation}
 \rho^{0(1)}_\lambda(r)=-\delta_{0(1)}\frac{d[\rho^n_{\rm g.s.}(r)\pm
 \rho^p_{\rm g.s.}(r)]}{dr}.
 \label{e6}
\end{equation}
It is straightforward to derive in this compact approach \cite{Kh03} a
consistent one-to-one correspondence between $\delta_{p(n)}$ and
$\delta_{0(1)}$, which can be used to determine $\delta_{0(1)}$ values from
$\delta_{p(n)}$ values given by the DWBA analysis. If one assumes that the
excitation is purely \emph{isoscalar} and the neutron and proton densities have
the same radial shape, scaled by the ratio $N/Z$, then
$\delta_n=\delta_p=\delta_0=\delta_1$. Therefore, any significant difference
between $\delta_0$ and $\delta_1$ would directly indicate a different isospin
distribution in the structure of the nuclear excitation under study.

In the present work, we have studied the elastic and inelastic $^{30,32}$S+$p$
scattering data at 53 MeV/u \cite{El01} and $^{34,36,38,40}$S+$p$ data
\cite{Al85,Ho90,Ke97,Ma99} at energies of 28 to 39 MeV/u. The IS and IV
contributions of the inelastic FF were considered explicitly to find out a
systematic behavior of $\delta_1$ along the Sulfur isotopic chain, passing by
the magic number $N=20$. Then, the folding + DWBA analysis of the elastic and
inelastic $^{18,20}$O+$p$ data at 43 MeV/u \cite{El00} and $^{22}$O+$p$ data at
46.6 MeV/u \cite{Be06} has been done to find out the $N$-dependence of
$\delta_1$ in the Oxygen case.

\begin{figure}[ht]
\centering\vspace{-0.5cm}\mbox{\epsfig{file=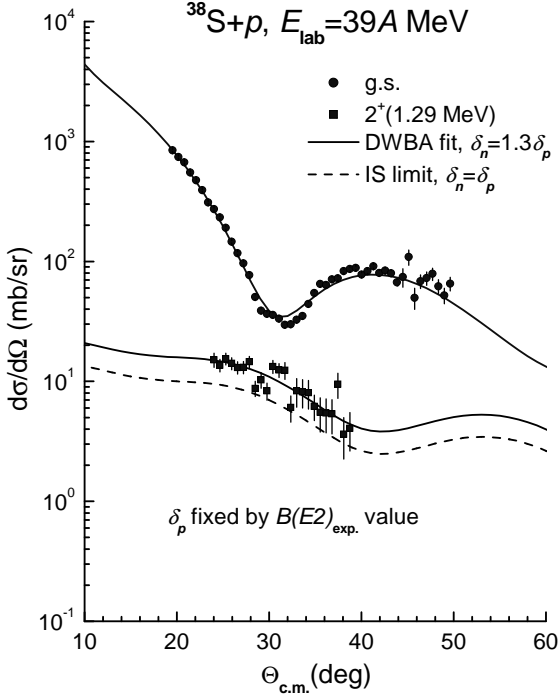,height=11.5cm}}
\vspace{-1cm}\caption{Elastic and inelastic $^{38}$S+$p$ scattering data at 39
MeV/u \cite{Ke97} in comparison with the DWBA cross sections given by the folded
FF. The dashed curve is inelastic cross section obtained in the isoscalar limit
($\delta_n=\delta_p=\delta_0=\delta_1$).} \label{f2}
\end{figure}

\begin{figure}[ht]
\centering\vspace{-0.5cm}\mbox{\epsfig{file=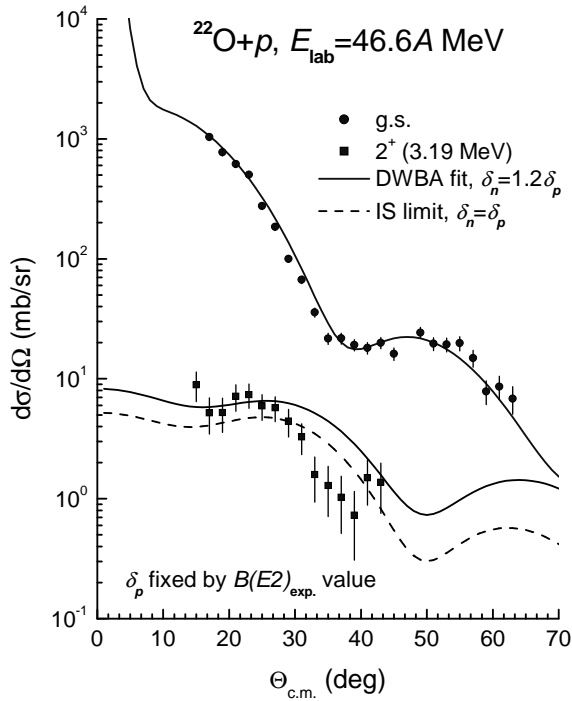,height=11.5cm}}
\vspace{-1cm}\caption{The same as Fig.~\ref{f2} but for the elastic and
 inelastic $^{22}$O+$p$ scattering data at 46.6 MeV/u \cite{Be06}.}
 \label{f4}\end{figure}

To have the accurate ``distorted" waves for the DWBA calculation of inelastic
scattering, the optical model (OM) analysis of the elastic data was done using
the real folded potential \cite{Kh02} obtained with the density- and isospin
dependent CDM3Y6 interaction \cite{Kh97} and nuclear g.s. densities given by the
Hartree-Fock-Bogoliubov (HFB) calculation \cite{Gr01}. The imaginary part of the
OP was taken in the Woods-Saxon (WS) form from the global systematics CH89
\cite{Va91}. All the considered elastic data were well reproduced (see
Figs.~\ref{f2} and \ref{f4}) with the depth of the WS imaginary potential
slightly adjusted by the OM fit (keeping the radius and diffuseness unchanged).
The experimental reduced electric transition rate $B(E2\uparrow)$ was used in
each case to fix $\delta_p$ value in the expression (\ref{e3}) for $\rho^p_2(r)$
which is further used in the folding calculation \cite{Kh03,Kh02}. As the only
parameter, $\delta_n$ was adjusted iteratively in the folding + DWBA calculation
to fit the measured inelastic cross section. Since the CDM3Y6 interaction is
real, the imaginary nuclear FF was obtained by deforming the imaginary part of
the OP with $\delta_0$ and $\delta_1$ values at each iteration step of the
folding + DWBA fit to the inelastic data. In each case, the final set of
deformation lengths $\delta_0$ and $\delta_1$ was fixed only after the best-fit
$\delta_n$ has been obtained.
\begin{figure}[ht]
\centering\vspace{-0.5cm}\mbox{\epsfig{file=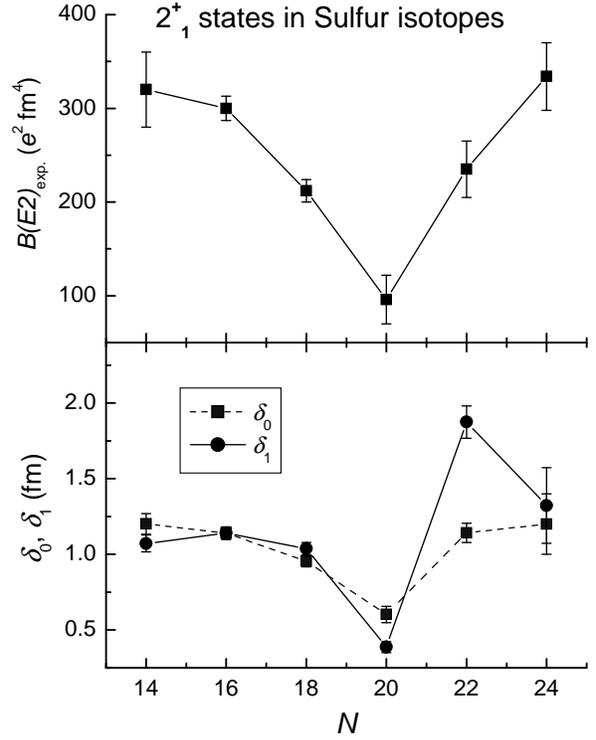,height=11.5cm}}
\vspace{-1cm}\caption{The measured $B(E2\uparrow)$ transition strength (upper
part), and the extracted isoscalar and isovector deformation lengths (lower
part) of 2$^+_1$ states in Sulfur isotopes. The lines are to guide the eye.}
 \label{f1}
\end{figure}

An earlier folding + DWBA analysis \cite{Kh02} of the same inelastic
$^{30-40}$S+$p$ scattering data, using inelastic FF given by the microscopic
transition densities obtained in the Quasiparticle Random Phase Approximation
(QRPA) \cite{El01}, has shown that the neutron and proton contributions to the
2$^+_1$ excitation in $^{30,32,34}$S follow approximately the \emph{isoscalar}
rule \cite{Kh03} which implies $\delta_0\approx\delta_1$. The present folding
model analysis using the collective-model transition densities (\ref{e3}) has
shown about the same results (see $\delta_0$ and $\delta_1$ values extracted for
$^{30,32,34}$S in Fig.~\ref{f1}). With the neutron shell becomes closed at
$N=20$, a significant ``damping" of the neutron transition strength occurs and
suppresses strongly the IV deformation length $\delta_1$ of the 2$^+_1$ state of
$^{36}$S. In fact, $\delta_1$ is reaching its minimum as the neutron number $N$
approaches the magic number 20. The $N$-dependence of the IV deformation length
$\delta_1$ is well correlated with the $N$-dependence of the reduced transition
rate $B(E2\uparrow)$ which also reaches its minimum at $N=20$ (see
Fig.~\ref{f1}). The shell closure effect is so strong in this case that the
proton and neutron QRPA transition densities needed to be scaled down by a
factor of 0.63 and 0.88, respectively, for a correct description of the measured
2$^+$ cross section in our earlier DWBA analysis \cite{Kh02} of the inelastic
$^{36}$S+$p$ scattering.
The contribution by the two valence neutrons in $^{38}$S to the 2$^+_1$
excitation is quite strong and $\delta_n$ turned out to be larger than
$\delta_p$ by around 30\% (see Fig.~\ref{f2}). This difference between the
proton and neutron transition strengths results on the IV deformation length
$\delta_1$ larger than $\delta_0$ by about 64\% (see Fig.~\ref{f1}). Note that
such a strong core polarization by the two valence neutrons in the $^{38}$S case
could not be fully accounted for by the QRPA calculation and the QRPA neutron
transition density has been scaled by a factor of 1.25 for a good agreement of
calculated DWBA cross section with the data \cite{Kh02}. Since the inelastic
$^{40}$S+$p$ scattering data at 30 MeV/u \cite{Ma99} consist of only two data
points with significant error bars, the extracted $\delta_{0(1)}$ values for
$^{40}$S might not be as reliable as those extracted for $^{30-38}$S. The new
measurement is, therefore, needed for an accurate estimate of the IV deformation
length of 2$^+_1$ state of $^{40}$S.

\begin{figure}[ht]
\centering\vspace{-0.5cm}\mbox{\epsfig{file=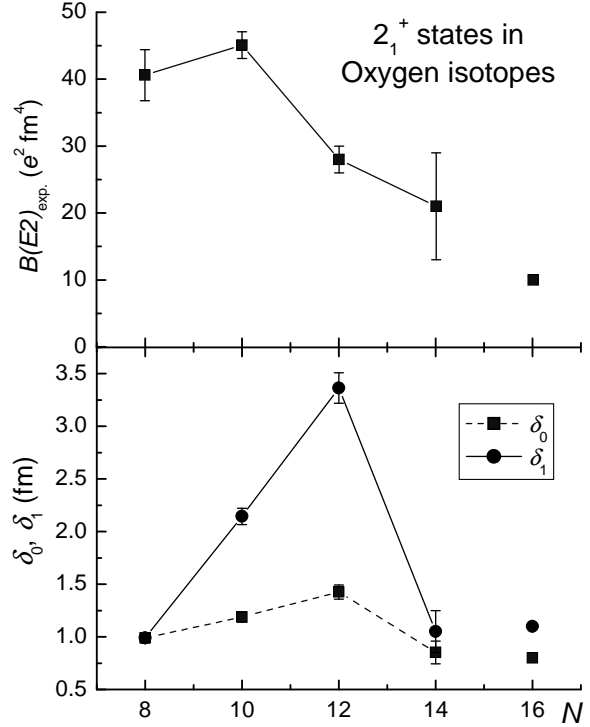,height=11.5cm}}
\vspace{-1cm}\caption{The measured $B(E2\uparrow)$ transition strength (upper
part) and the extracted IS and IV deformation lengths (lower part) of 2$^+_1$
states in the Oxygen isotopes. The values for $^{24}$O are deduced from the QRPA
prediction by Khan {\it et al.} \cite{El02}. The lines are to guide the eye.}
\label{f3}
\end{figure}

The present analysis of the inelastic $^{18,20,22}$O+$p$ scattering data has
shown an interesting $N$-dependence of $\delta_1$. Except for the double-magic
($N=Z$) $^{16}$O, where one has exact relation $\delta_0=\delta_1$, the results
obtained for $^{18}$O already indicate a rather strong IV deformation for the
2$^+_1$ state of this nucleus (see Fig.~\ref{f3}). We recall that the IS and IV
deformation parameters of the 2$^+_1$ state of $^{18}$O have been determined
long ago by Grabmayr {\it et al.} \cite{Gr80} in a simultaneous DWBA analysis of
the $(p,p')$ and $(n,n')$ inelastic scattering data at 24 MeV using the
collective form factor (\ref{e2}). It is easy to deduce from the results of
Ref.~\cite{Gr80} the corresponding deformation lengths $\delta_0\approx 1.1$ and
$\delta_1\approx 2.6\pm 1.3$ fm for the 2$^+_1$ state of $^{18}$O. These values
agree reasonably with the results of our folding + DWBA analysis (see
Fig.~\ref{f3}) of the inelastic $^{18}$O+$p$ scattering data at 43 MeV/u
\cite{El00}. Note that the $\delta_1$ value extracted from our analysis has a
much smaller uncertainty  compared to that deduced from the results of
Ref.~\cite{Gr80}. The error bars for $\delta_{0(1)}$ plotted in Figs.~\ref{f1}
and \ref{f3} were accumulated from the experimental uncertainties of the
measured $B(E2)$ values and $(p,p')$ cross section.

As already mentioned above, a double measurement of $(p,p')$ and $(n,n')$
inelastic scattering is not possible with unstable Oxygen isotopes, and our
folding method is the only alternative way to determine $\delta_1$. The present
analysis (with a more consistent treatment of the IV part of the imaginary FF)
has confirmed again the large IV deformation length of the 2$^+_1$ state of
$^{20}$O found earlier in Ref.~\cite{Kh03}. With the IV deformation about three
times the IS deformation (see Fig.~\ref{f3}), the contribution by the Lane form
factor $F_1$ to the 2$^+_1$ excitation of $^{20}$O amounts up to 40-50\% of the
total inelastic cross section \cite{Kh03}. If we consider $^{20}$O as consisting
of the $^{16}$O (or $^{18}$O) core and four (or two) valence neutrons, then a
large value of IV deformation length $\delta_1$ indicates a strong core
polarization by the valence neutrons in the 2$^+_1$ state of $^{20}$O.

In such a ``core + valence neutrons" picture, it is natural to expect that the
2$^+_1$ state of $^{22}$O should be more collective and have a larger IV
deformation length due to the contribution of two more valence neutrons.
However, the inelastic $^{22}$O+$p$ scattering data at 46.6 MeV/u measured
recently at GANIL \cite{Be06} show clearly the opposite effect, with the
$(p,p')$ cross section about 3 to 4 times smaller than that measured for the
$2^+_1$ state of $^{20}$O at 43 MeV \cite{El00} over a wide angular range. The
folding + DWBA analysis \cite{Be06} of these data using the QRPA transition
densities for the $2^+_1$ state of $^{22}$O has pointed to a much weaker neutron
transition strength compared to that of the $2^+_1$ state of $^{20}$O. Given a
significantly higher excitation energy of this state (1.5 MeV higher than that
of the $2^+_1$ state of $^{20}$O), the newly measured inelastic $^{22}$O+$p$
scattering data were suggested \cite{Be06} as an important evidence for the
neutron shell closure at $N=14$ or 16.

With the proton deformation length $\delta_p$ fixed by the adopted
$B(E2\uparrow)$ value for the $2^+_1$ state of $^{22}$O \cite{Ra01}, the present
analysis has given the \emph{upper limit} for the neutron deformation length as
$\delta_n\leq 1.2~\delta_p$ (see Fig.~\ref{f4}), which implies $\delta_1\leq
1.1$ fm. As a result, the extracted $\delta_1$ value is just slightly larger
than $\delta_0$ and the $N$-dependence shown in Fig.~\ref{f4} indicates that
$\delta_1$ reaches its minimum at either $N=14$ or 16. Based on the results
obtained above for the Sulfur isotopes, we conclude that the deduced
$N$-dependence of $\delta_1$ for the $2^+_1$ states of Oxygen isotopes suggests
the neutron shell closure at either $N=14$ or 16. We note further that the
$B(E2\uparrow)$ value predicted by the QRPA calculation \cite{El02} for the
$2^+_1$ state of $^{24}$O is even smaller than that adopted for the $2^+_1$
state of $^{22}$O, a fact which could favor the shell closure at $N=16$. We have
also deduced $\delta_0$ and $\delta_1$ of the $2^+_1$ state of $^{24}$O from
these QRPA results and they are nearly the same as those deduced from the
$(p,p')$ data in $^{22}$O case. If we take into consideration 4 MeV gap between
the 2s$_{1/2}$ and 1d$_{3/2}$ subshells predicted recently by a consistent HFB
calculation \cite{Ob05}, then it is also likely that the neutron shell closure
occurs at $N=16$. In any case, more measurements for $^{24}$O are highly
desirable for a definitive conclusion on a new magic number $N=16$ in the
neutron rich nuclei.

In summary, we have shown that the behavior of the dynamic isovector deformation
of the $2^+_1$ states in neutron rich nuclei is closely correlated with the
evolution of the valence neutron shell. This interesting result emphasizes again
the importance of $(p,p')$ reactions measured with unstable nuclei in the
inverse kinematics.

\section*{Acknowledgement}
This research project has been supported, in part, by the Natural Science
Council of Vietnam, EU Asia-Link Program CN/Asia-Link/008 (94791) and Vietnam
Atomic Energy Commission (VAEC).

\end{document}